\documentclass[a4paper]{article}
\usepackage{interspeech2010,amssymb,amsmath,epsfig}
\usepackage[safe]{tipa}
\usepackage{tipx,nicefrac,url}
\ninept
\def\reg{{\rm\ooalign{\hfil
     \raise.07ex\hbox{\scriptsize R}\hfil\crcr\mathhexbox20D}}}

\title{Detecting gross alignment errors in the Spoken British National Corpus}

\makeatletter
\def\name#1{\gdef\@name{#1\\}}
\makeatother
\name{{\em Ladan Baghai-Ravary, Sergio Grau, Greg Kochanski}}

\address{Phonetics Laboratory, University of Oxford, United Kingdom \\
{\small \tt \{ladan.baghai-ravary,sergio.graupuerto,greg.kochanski\}@phon.ox.ac.uk}}

\begin{document}
\maketitle
\begin{abstract}
The paper presents methods for evaluating the accuracy of alignments between transcriptions and audio recordings. The methods have been applied to the Spoken British National Corpus, which is an extensive and varied corpus of natural unscripted speech.
Early results show good agreement with human ratings of alignment accuracy. The methods also provide an indication of the location of likely alignment problems; this should allow efficient manual examination of large corpora.
Automatic checking of such alignments is crucial when analysing any very large corpus, since even the best current speech alignment systems will occasionally make serious errors. The methods described here use a hybrid approach based on statistics of the speech signal itself, statistics of the labels being evaluated, and statistics linking the two.
\end{abstract}

\section{Introduction}
In linguistic and phonetic research, increasing emphasis is being placed on the analysis of very large corpora, to give more general, stronger conclusions. However, the analysis of such databases often requires aligned textual and audio data, and performing such alignments manually is extremely time-consuming and hence expensive.
Automatic alignment can be performed by transcribing the words spoken in the corpus and then aligning them to the speech via standard HMM techniques (e.g. \cite{Sjolander}). These
Automatic Speech-to-Phoneme Alignment (ASPA) systems can produce accurate estimates of word and phoneme locations, but often fail in cases such as:
\begin{enumerate}
\item The speech has been recorded in an environment with non-stationary background noise, competing un-transcribed speech, distortion, and/or reverberation.
\item The phonemic transcription of the speech is not accurate.  The phonemic transcription is usually obtained by a dictionary look-up process. Citation forms are often used, and are often unrealistic, especially when dealing with spontaneous speech.  In natural, unscripted speech, people will sometimes talk simultaneously so there may be no sequence of words or phonemes that can correctly represent the audio. Even if the words can be identified and transcribed individually, they cannot be organised into a simple sequence, and so require more complex techniques (e.g. \cite{pmc}), which are impractical for large corpora.
\item The word-level transcription is inaccurate and/or inconsistent in its handling of nonspeech sounds, backchannels, mumbles, and speech-like noises (e.g.\ dog barks).
\item The speech is only available in long continuous recordings: the accumulation of HMM probabilities over extended periods of time can introduces numerical errors which distort the alignment process \cite{Toth}.
\end{enumerate}

As a result, automatic alignment of large quantities of spontaneous unconstrained speech is invariably error-prone.
When labelling ``speech in the wild'' such as the Spoken British National Corpus (BNC) \cite{BNC}, the above four conditions frequently occur, and lead to failures of the alignment system. Identifying where such an alignment succeeds and where it fails allows bad regions to be avoided or realigned.

Previous work on this topic is sparse. 
Traditionally, aligner accuracy is assessed by comparison with manually estimated labels in one way or another \cite{Villiers2006}. However, the task here is fundamentally different in that we operate under conditions where any aligner may fail, regardless of whether it is generally accurate or not.
\cite{Barnard2006} have addressed our problem in a different context (clean speech that is intended for use in TTS systems) with some success.
They flagged 24\% of the segments as suspicious and detected 43\% of the total errors, which would have led to a modest reduction in the effort required to verify a corpus.  \cite{Davel2005} looked at the related problem of finding errors in the lexicon used in the alignment process.  Some text-to-speech system builders may also have used similar ideas (e.g.\ \cite{Whistler}) but details are lacking.  Some preliminary work was done by \cite{Das2010}, using ideas related to our \textit{improbable} and \textit{unexpected} features.

The work described here is part of a project \cite{Coleman2011} to label speech from the Spoken BNC, originally recorded on analogue cassette tapes between 1991 and 1994. The data consists of recently-digitised recordings with an associated word-level transcription of the audio data. The recordings are mostly of unscripted, spontaneous speech, and include a diverse range of recording conditions, accents, and microphone positions relative to the speakers. Each track of the original cassette recordings has been digitised to a single file, so the data to be aligned is generally just over either 45 or 60 minutes long. Background noise varies widely, both in terms of amplitude and character (competing speech, mechanical noise, music and speech from television, radio or other sources, microphone-handling noise, etc.).

Our earlier work \cite{Ravary:2009,Ravary:2009a} investigated alignment errors by comparing the alignments produced by a large number of ASPA systems. However, in that work we were attempting to assess the general ability of aligners to identify particular phone-transitions, and the general quality of alignments produced by a specific system, respectively.

\section{Methods}

We used human evaluations of overall alignment quality for each of a set of recordings, to construct algorithms that should be able to identify suspicious regions, then to compare these to the human evaluations.

\subsection{Alignment Procedures}
To evaluate our methods, we used the Penn Phonetics Lab Forced Aligner, P2FA\cite{Yuan2008} to analyse 46 recordings taken from the BNC, each consisting of a full audio session recorded on one side of an audio tape without a break.

P2FA is an automatic phonetic aligner based on HTK, and developed at the Phonetics Laboratory of the University of Pennsylvania. It employs monophone Gaussian Mixture Model based HMMs which were trained using 39 perceptual linear prediction (PLP) coefficients. 
We used the 39 phone set from the Carnegie-Mellon University Pronouncing Dictionary, CMUdict \cite{cmudict}, with the addition of OH for British English /\textturnscripta/ distinct from /\textinvscripta/, including lexical stress marking for the vowels. The current CMU Pronouncing Dictionary was extended to include all the out-of-vocabulary words and to include a range of common British English word pronunciations. This extension was performed using semi-automatic methods by experienced phoneticians.

\subsection{Human Evaluations}
We observed that some of the alignments were much more successful than others, and we quantified that with a rating procedure.
One of the authors examined 5-second long regions, approximately once every 60 seconds throughout the files, and checked whether that region was correctly aligned at a word level.
The overall score of the file was subjective on a scale between 0 (very poor) and 10 (very good), but was intended to reflect the number and magnitude of alignment problems.
\subsection{Algorithms}
We developed five algorithms to indicate potential problems with an alignment, listed below.
Each one takes the aligner output (segment label times, and log-probabilities) and optionally the audio file, and identifies a list of suspicious locations. 
In these descriptions, $L_i$ is the aligner's HMM log-probability value for phoneme instance $i$, $p_i$ is the phoneme (i.e.\ /a/, /t/, /\textesh
/, ...), and $\delta_i$ is the duration.  The algorithms were developed without reference to the human evaluations, except for the setting of each algorithm's threshold.

\textbf{Unexpected Log(P)}: This algorithm builds a prediction of the aligner's log-probability score per unit length from the corpus as a whole (except the data file under analysis).  It then computes the difference between $L_i/\delta_i$ and the prediction\footnote{We drop phones with $\delta_i =0$ or with $\delta_i \ge 1s$ on the grounds that the scaling of $L_i$ with $\delta_i$ may not be accurate on these phones.}.
It operates on the assumption that most of the audio in the corpus is correctly aligned so that its predictions correspond to good alignment.   Thus, when the aligner is doing worse than usual, and $L_i$ is low, the difference will be substantially negative.
We have observed that when the aligner fails, it typically fails for a relatively large region: a word or more.  To make use of this knowledge, we smooth the difference over a 1~second long region.  When this smoothed difference of $log(P)$ is more negative than a threshold, the algorithm has identified a suspicious region.

The predictor for $L_i$ starts with the median value for that phoneme, $\lambda(p_i) = \mathrm{median}(L_j / \delta_j \; \mathrm{if} \; p_j=p_i)$.
It then adds in a 5-term linear prediction.  The independent variable in the first term is $log(\delta_i/D(p_i))$, where $D$ is the median duration of a phoneme class, and $D(p_i) = \mathrm{median}( \delta_j \;\mathrm{if}\; p_j=p_i )$.
The remaining four terms capture some information on the phoneme sequence.
The second captures the typical difference in duration between the phone class under consideration and the previous phone class: $log(D(p_i)/D(p_{i-1}))$.  The third captures the typical difference in $log(P)$ between the phoneme class under consideration and the preceding phoneme class: $\lambda(p_i)-\lambda(p_{i-1})$.  The fourth and fifth are the same, except they refer to the succeeding phoneme.
The predictor (along with the medians) is trained on phonemes with $0.04s \le \delta_i \le 0.18s$.
The output of this algorithm becomes the ``unexpected'' feature.

\textbf{Word Log(P)}: we consider words with four or more phonemes to avoid variations due to the vagaries of individual phonemes. The individual $L_i$ values for each phoneme are summed over the respective word, then normalised by dividing by the word duration. This log probability per unit time provides a stable indication of the goodness of fit of the observed data to the HMM. The final stage of of the \textit{Word Log(P)} method compares the normalised log probability with a fixed threshold, yielding the ``improbable'' feature. This  method is complemementary to the \textit{Unexpected Log(P)} method, above, in that it combines data from a whole word, whereas the \textit{Unexpected Log(P)} method utilises features at the individual phoneme level, before smoothing them, i.e.\ the \textit{Unexpected Log(P)} algorithm normalises $L_i$ by duration over a larger unit.

In the \textit{Unexpected Log(P)} algorithm, the threshold was set by experiment, to identify about 100 suspicious regions per hour on files that had substantial alignment problems. The \textit{Word Log(P)} threshold was set to identify a similar number of events on poorly aligned files, and typically fewer on well-aligned ones. 

\textbf{Extremes of Amplitude}: Many alignment systems will produce erroneous alignments when several consecutive speech labels bunch-up into a short region, with the remaining speech labelled as an extended silence. This misidentification of speech and silence can be detected by examining the amplitude of signals in each labelled phoneme. A contiguous region of quiet, of a length comparable to a short word within a segment labelled as speech, indicates that an error may have occurred. Similarly, an error is likely if there is a word-length region of high amplitude within a segment labelled as silence. High amplitude ``silence'' regions should be marked for human inspection even if they do not contain speech, because they represent high-amplitude background noise, which is itself a potential cause of problems in real-world data.

The thresholds for ``quiet'' and ``high amplitude'' were set as the $3^\mathrm{rd}$ and the $97^\mathrm{th}$ percentile of the amplitudes observed over the whole of the recording. The nominal length of a ``short word'' was set to $\nicefrac{1}{4}$~second, assuming four phonemes with an average duration of $\nicefrac{1}{16}$~second each. These parameters were estimated by experiment, and chosen to give a relatively small number of false positives.  This algorithm produces two factors (``loud'', and ``quiet'') as it reports the extremes separately.

\textbf{Word Duration}: this algorithm simply examines the durations of the segments, and if there are any which are unexpectedly long or short, indicates an error. It is difficult (simply from their duration) to detect periods of silence which have become extended or merged due to an alignment error. But the durations of segments labelled as speech can be of great help.

This algorithm takes a word-based approach to detecting unusual segment durations.
Individual phoneme durations are not reliable indicators because of variabilities in pronunciation due either to dialect, style of speaking, or the effects of transient background noise.
Thus we take all words with four or more phonemes, calculate the duration of the region labelled as the word, normalise it by dividing by the number of phonemes, and compare it with two thresholds representing the largest and the smallest average phoneme duration. Any result outside the range $\nicefrac{1}{32}\;\mathrm{s} < \mbox{mean duration} < \nicefrac{1}{8}\;\mathrm{s}$ is flagged. The lower threshold is just above the minimum possible duration of a phoneme label for our HMMs (which use 3 left-to-right states per phone).
This algorithm yields two features (``short'' and ``long''), as it reports the two extremes separately.

\textbf{Duration Mismatch}:  This algorithm builds a duration model for phonemes and then measures how far each phoneme\footnote{We do not compute a result for phonemes with $\delta_i =0$ or with $\delta_i \ge 1s$ on the grounds that they are outside the range of validity of the duration model, and are almost all silences, anyway. However, these phones may be used as neighbours in the computation of other phones.} deviates from the model.
Regions are identified as suspicious if the smoothed absolute value of the deviation is large enough to exceed a threshold.

The duration model predicts the log of the phoneme duration as $d_i$. It starts with a value typical of that phone: $\Delta_i = \mathrm{median}(\log(\delta_j) \;\mathrm{if}\; p_i=p_j)$.  It then adds on a 25-term linear predictor: the constant term captures a constant offset from $\Delta_i$.
Then, twelve terms capture the length of nearby phonemes relative to their median durations (six neighbours on each side), via factors that are $\mathrm{q}(D_i \delta_{i+k}, D_{i+k} \delta_i ),$ where k specifies which neighbour\footnote{In practice, these first 12 terms amount to a normalisation of the duration for changes in the local speech rate: the model adjusts the phone duration by 36\% of the average change in nearby durations}.
The $\mathrm{q}(a, b) = \begin{Bmatrix} 2(a/b)^{0.5} -2, \;\mathrm{if}\; a\le b \\ 2-\mathrm{q}(b,a), \;\mathrm{else} \end{Bmatrix}$ function is a sigmoid whose domain is $[0, \infty]$, and it is well-behaved at the endpoints, an important property since some phoneme durations are zero.  The final 12 terms similarly capture the differences between the typical durations of neighbouring phones.  The are represented by features $\mathrm{q}(D_i, D_{i+k})$.  This duration model is trained to match $\mathrm{log}(\delta_i)$ as in the \textit{Unexpected Log(P)} algorithm.

Finally, each phone is scored by $S_i = | \delta_i - d_i | / m_i$, where $m_i = \mathrm{median}( | \log(\delta_j)-\Delta_i | \;\mathrm{if}\; p_j = p_i )$ is the mean absolute deviation of the log duration.  The scores are then smoothed and thresholded as in the \textit{Unexpected Log(P)} algorithm.  This produces the ``badlength'' feature.

\section{Results}

\begin{figure}
\centerline{\epsfig{figure=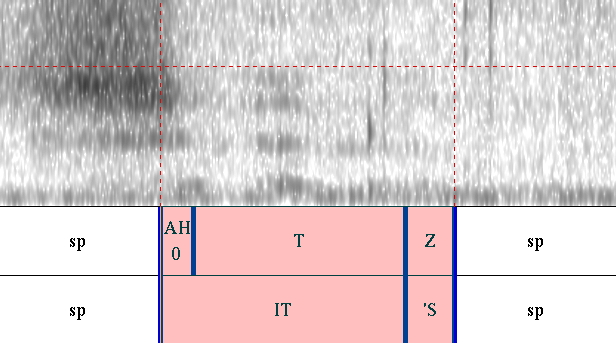, height=3.9cm}}
\caption{{\it Quiet segment error: the ``quiet'' detector has identified a region (shaded), labelled as the word ``it's''. The vowel is omitted from the labelled region, and the end of the region extended into silence. The whole region is very quiet.  The top tier is the spectrogram, then phoneme and word labels, respectively.}}  
\label{fig:exampleQuiet}
\end{figure}

\begin{figure*}
\centerline{\epsfig{figure=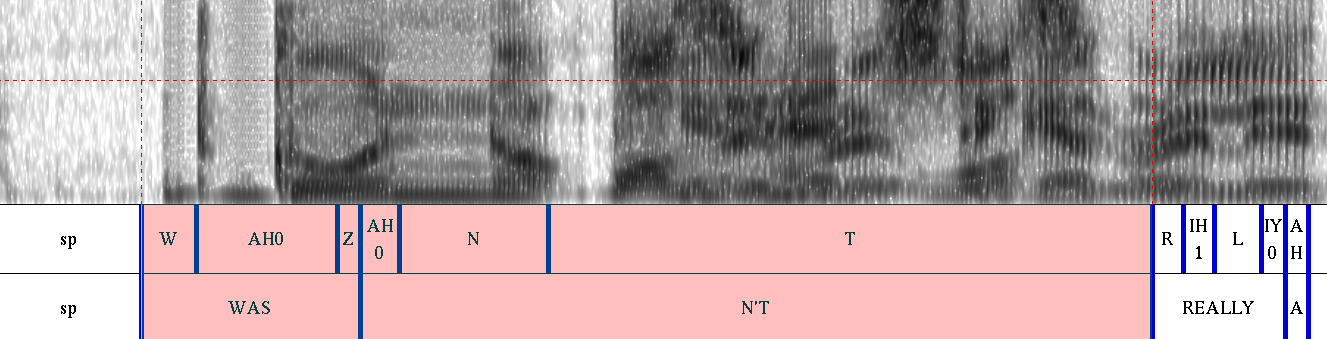, height=3.9cm}}
\caption{{\it Long word error. Displayed as per Figure~\ref{fig:exampleQuiet}, it shows where the ``long'' detector has identified a region (shaded in the Figure) that was aligned as a single word, ``wasn't'', but actually included several words.}}  
\label{fig:exampleLong}
\end{figure*}

Figures~\ref{fig:exampleQuiet} and~\ref{fig:exampleLong} show two examples of regions correctly identified as misaligned. Many such identifications are correct. We combined the results for each audio file to give an overall score based on the total number or duration of suspected of regions.
The outputs of the above seven features were then correlated with the evaluations using a linear regression, via the \textit{``glm''} method of the R software package \cite{R}.

The lengths of the audio files varied, as did the amount of speech, so we devised three ways to define a score, and since we had no clear criterion to pick one over the other, we computed separate linear regressions with each.  The first, $s^{nd}$, is the number of identified regions divided by the duration of the audio file; the second, $s^{nw}$, is the number of regions divided by the number of words in the audio file (as determined from the BNC transcriptions); and the third, $s^{dd}$, is the total duration of the regions divided by the duration of the audio file.

The distributions of the evaluation variable and the various $s$ variables were strongly non-Gaussian, with a maxima at one edge.
Transforming the independent variables by raising them to the power 0.3 made the distribution subjectively more normal, as did squaring the evaluations. However, these transforms were not more than partially successful, so we elected to regress both with and without the transforms.
This led to four regressions for each choice of $s$, or 12 regressions in all.

Of those 12 regressions, \textit{short} was statistically significant on 10 ($P<0.01$), \textit{badlength} was significant on 6, \textit{loud} on 4.\footnote{Note that with 12 regressions, we expect 3 false significances at the $P<0.05$ level and one at the $P<0.01$ level. Therefore, of the 9 significances reported at the $P<0.05$ level (4 for \textit{badlength}, 3 for \textit{loud}, 2 for \textit{long}), half are probably spurious. For simplicity, we will ignore them all.} At least one of those factors was significant at the $P<0.01$ level in each regression.  Pearson's $R^2$ averaged 0.66 with a standard deviation of 0.13 over the regressions, indicating that a combination of the algorithms was reasonably effective at matching the human judgements of overall alignment accuracy.

Of the best three fits ($R^2=0.81$, $0.81$, $0.87$), \textit{short} was significant on each at $P<0.01$, along with \textit{badlength} and \textit{loud} once each.
Of these, one used $s^{dd}$ and did not transform the data at all; the other two used $s^{nw}$ and did not transform the independent variables.

\begin{figure*}
\centerline{\epsfig{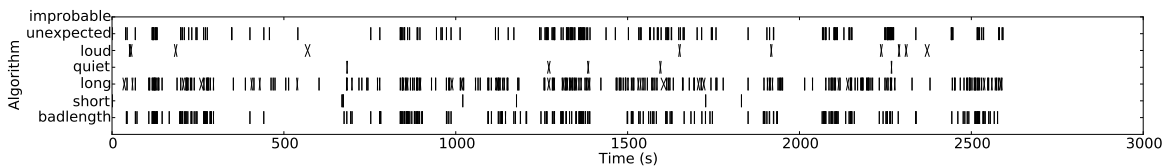}}
\caption{{\it Identified regions: each row shows where an algorithm flagged a potential alignment error.  The audio file was rated 8 for overall alignment success.
NB: the \textit{improbable} detector never fired on this file. 
}}  
\label{fig:correlation}

\end{figure*}

In the second analysis, the correlations between the various algorithms were calculated for each recording.
Figure \ref{fig:correlation} shows the results for one recording.
Each pair of algorithms scored a point when they identified a pair of regions within 5~s of each other. We compared the number of such pairs to the number of accidental pairs that would occur if the algorithm's outputs were uncorrelated with each other. 

There is a pattern of correlation between some of the detectors in Figure~\ref{fig:correlation}. Several pairs of algorithms were strongly correlated: notably \textit{badlength} and \textit{unexpected}, \textit{badlength} and \textit{long}, and \textit{long} and \textit{unexpected}.  These coincided 3.6 to 5.1 times more often than chance, with statististical significances well beyond $P < 0.001$.
Several of the pairs of algorithms were anticorrelated, notably \textit{loud} vs.\ \textit{short}, \textit{loud} vs.\ \textit{badlength}, and \textit{loud} vs.\ \textit{long}.  This is due to the designs of the algorithms: specifically, \textit{loud} triggers only on silences, while the others trigger only on speech sounds.
As a result, they never pick the same phoneme, and only occasionally pick phonemes within 5~s of each other.
The remaining pairs were either nearly independent (\textit{loud} \& \textit{unexpected}, \textit{badlength} \& \textit{short}, \textit{long} \& \textit{short}, \textit{short} \& \textit{unexpected}) or did not have enough occurrences to draw any reliable conclusion.

Duration-based measurements seem to perform best (i.e.\ \textit{short} and \textit{badlength}).
One of the most useful indicators of a gross alignment error was also the simplest: the \textit{short} algorithm detected a sequence of phonemes whose durations were at the minimum allowed by their state topology (here, 3~states or 30~milliseconds).
The relative lack of success of the \textit{improbable} and \textit{unexpected} algorithms was unexpected: good and bad alignments have similar distributions of Log(P) scores.

\section{Conclusions}
We have shown that automated techniques can usefully identify regions of bad alignment.
The regions identified by some of our algorithms correlate well with human evaluations of the overall quality of the alignment.
In general it appears that the most reliable method for judging the quality is simply to consider the statistics of the segment durations, either over a fixed time window, or a linguistic unit (e.g. a word).
This research should allow semi-automatic evaluation of the alignment of large speech corpora, which will be important for their future use in speech research.

\section{Acknowledgements}
We thank JISC (in the UK) and NSF (in the USA) for their support of \textit{Mining a Year of Speech\/}, under the \textit{Digging into Data} programme. This work is also partly supported by the UK ESRC (awards RES-062-23-2566, RES-062-23-1172, and RES-062-23-1323).  We thank John Coleman and Ranjan Sen for the dictionaries, and John Coleman for his comments.

\eightpt
\bibliographystyle{IEEEtran}

\end{document}